# Single and Double Scattering Mechanisms in Ionization of Helium by Electron Vortex Projectiles

A. L. Harris*

Physics Department, Illinois State University, Normal, IL, USA 61790

## Abstract

Triple differential cross sections (TDCSs) for electron vortex projectile ionization of helium into the azimuthal plane are calculated using the distorted wave Born approximation. In this collision geometry, the TDCSs at low and intermediate energies exhibit unique qualitative features that can be used to identify single and double scattering mechanisms. In general, our results predict that the ionization dynamics for vortex projectiles are similar to those of their non-vortex counterparts. However, some key differences are observed. For non-vortex projectiles, a double scattering mechanism is required to emit electrons into the azimuthal plane, and this mechanism becomes more important with increasing energy. Our results demonstrate that for vortex projectiles, emission into the azimuthal plane does not require a double scattering mechanism, although this process still significantly influences the shape of the TDCS at higher energies. At low projectile energies, non-vortex ionization proceeds primarily through single binary collisions. The same is generally true for vortex projectiles, although our results indicate that double scattering is also important, even at low energy. Vortex projectiles have an inherent uncertainty in their incident momentum, which causes a broadening of the binary peak at all energies and results in a splitting of the binary peak at higher energies. The results presented here lead to several predictions that can be experimentally tested.

## 1. Introduction

Triple differential cross sections (TDCSs) for ionization processes have long been used to study the mechanisms that lead to electron removal from atoms and molecules. Many studies focus exclusively on the coplanar scattering geometry in which the incident, scattered, and ionized electrons are all observed in the same scattering plane. However, much richer scattering dynamics can be studied by examining collisions in which the final state free electrons are observed outside the scattering plane. In particular, when the final state electrons are observed in a plane perpendicular to the incident beam direction (azimuthal plane), double and single scattering mechanisms can be isolated [1–7]. For plane wave electron projectiles at low energy and symmetric energy sharing, a single binary peak is observed in the azimuthal plane TDCS, while at higher energies, two peaks are observed [1–3]. These qualitatively different structures in the TDCS are due to different scattering mechanisms, and simple classical descriptions can be used to explain these features [1,2].

* alharri@ilstu.edu

For a binary collision with the target electron at rest, ejection of the atomic electron into the azimuthal plane is forbidden due to momentum conservation. However, the inclusion of the quantum mechanical target electron momentum density leads to a non-zero probability of finding the electron in the azimuthal plane due to a binary collision. Because the target electron momentum density for ground state atoms is strongly peaked near zero, the outgoing electrons are most likely to be emitted in a back-to-back geometry. This results in the single peak structure observed at low energies. However, other emission geometries are possible with more complex scattering mechanisms, such as the double scattering that leads to perpendicular emission of the electrons, as demonstrated in [1,2]. Therefore, the qualitative features of the azimuthal plane TDCS can be used to identify scattering mechanisms.

More recently, the creation and production of sculpted electron wave packets, such as Bessel or Airy electrons, has spurred interest in collisions between non-plane wave projectiles and atoms and molecules [8–21]. These sculpted electrons are quite different from their plane wave counterparts because they can carry quantized angular momentum and have non-zero transverse linear momentum. Previous studies of ionization by sculpted electrons have shown that the TDCSs for ionization by vortex projectiles are qualitatively and quantitatively different from those of non-vortex projectiles [8–10]. It is therefore natural to ask whether the ionization mechanisms for vortex projectiles are different than those of their non-vortex counterparts.

Because the qualitative shape of the azimuthal plane TDCS provides clues as to the ionization mechanism, this collision geometry is ideal for studying vortex projectile ionization. We present TDCSs for ionization of helium using electron vortex (EV) Bessel projectiles. Specifically, we examine TDCSs for ionization into the azimuthal plane and show that at low energies, ionization by vortex projectile proceeds primarily through binary collisions, as in the case of non-vortex projectiles. However, our results predict that even at low energy, the double scattering mechanism alters the magnitude of the TDCS. For higher energy, inclusion of the double collision mechanism in the model alters the shape of the TDCS and leads to a predicted enhancement of electron emission in the azimuthal plane geometry. Atomic units are used throughout.

## 2. Theory

### 2.1 Collision Geometry

The collision geometry is shown in Fig. 1, with the collision occurring at the origin. The incident projectile propagates in the z-direction and without loss of generality is scattered along the positive x-axis following the collision. The x-z plane is the scattering plane and the azimuthal plane is the x-y plane. In the final state, both electrons are observed in the azimuthal plane with a relative angle between them of $\varphi_e$. An ionized electron with azimuthal angle $\varphi_e = 180°$ or $0°$ is common to both the scattering plane and azimuthal plane. The final state electrons have equal energy, as found from energy conservation

$$E_f = E_e = \frac{E_i - I_p}{2},$$

where $E_{f,e}$ is the energy of the scattered (ionized) electron, $E_i$ is the incident projectile energy, and $I_p$ is the ionization potential of the target atom (24.6 eV for helium). As detailed below, for a vortex projectile, the incident momentum direction is uncertain, but lies along a cone (purple arrows in Fig. 1) centered on the z-axis with half angle $\alpha$, which is referred to as the vortex opening angle.

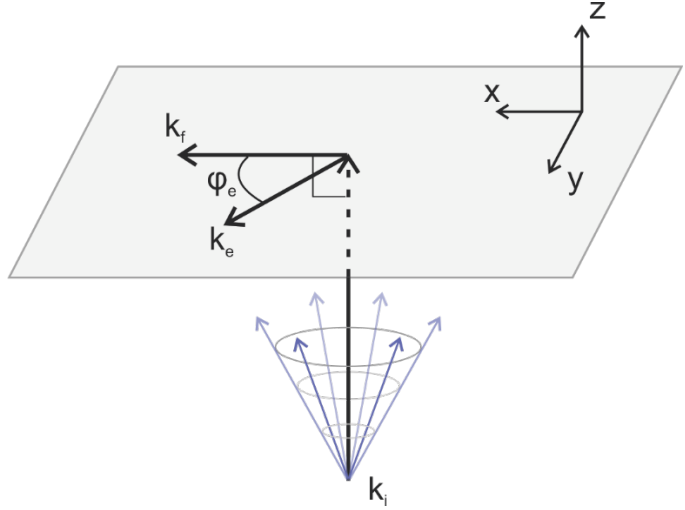

Figure 1 Collision geometry for (e,2e) ionization into the azimuthal (x-y) plane. The incident projectile propagates in the z-direction, with the incident vortex momentum vectors lying on a cone of half angle $\alpha$ (purple vectors). The scattered projectile momentum $\vec{k}_f$ defines the x-axis. The ionized electron's momentum $\vec{k}_e$ lies in the azimuthal plane at an angle $\varphi_e$ relative to the scattered projectile.

## 2.2 Triple Differential Cross Sections

For ionization by EV beam, the TDCS within a first order perturbative model is given by [8,10]

$$\frac{d^3\sigma}{d\Omega_1 d\Omega_2 dE_2} = \mu_{pa}^2 \mu_{ie} \frac{k_f k_e}{k_i} \left| T_{fi}^V(\vec{q}) \right|^2, \qquad (1)$$

where $T_{fi}^V(\vec{q})$ is the transition matrix element, and the momenta of the incident projectile, scattered projectile, and ionized electron are $\vec{k}_i, \vec{k}_f, \vec{k}_e$ respectively. A key parameter for calculation of the TDCS is the momentum transfer vector, which is defined by the projectile's change in momentum $\vec{q} = \vec{k}_i - \vec{k}_f$. The reduced masses of the projectile and target atom and the He$^+$ ion and ionized electron are $\mu_{pa}$ and $\mu_{ie}$, respectively. For symmetric energy sharing, the

TDCS of Eq. (1) must be multiplied by 2 because the indistinguishability of the projectile and target electrons results in the equality of the direct and exchange amplitudes contained in $T_{fi}^V(\vec{q})$.

Explicitly using the direct term in the transition matrix,

$$T_{fi}^V = -(2\pi)^{\frac{3}{2}} < \Psi_f|V|\Psi_i^V >, \tag{2}$$

where $\Psi_i^V$ is the initial state vortex wave function, $\Psi_f$ is the final state non-vortex wave function, and $V$ is the perturbation. Unlike the traditional plane wave projectile, vortex projectiles are not uniform in the transverse direction. In the case of the Bessel wave function used here, the incident projectile has a phase singularity at a particular spatial location and an impact parameter $\vec{b}$ can be used to describe the transverse location of the phase singularity relative to the z-axis. Calculation of the transition matrix of Eq. (2) then occurs for a specific impact parameter. However, in a traditional collision experiment using a gas phase target, the impact parameter cannot be controlled and comparison with theory requires an average over impact parameters. This average leads to an expression for the TDCS that can be written in terms of the plane wave transition matrix element $T_{fi}^{PW}(\vec{q}) = -(2\pi)^2 < \Psi_f|V|\Psi_i^{PW} >$, such that [8,10,18,19]

$$\frac{d^3\sigma}{d\Omega_1 d\Omega_2 dE_2} = \mu_{pa}^2 \mu_{pi} \frac{k_f k_e}{(2\pi) k_{i_z}} \int d\phi_{k_i} \left|T_{fi}^{PW}(\vec{q})\right|^2, \tag{3}$$

where $\phi_{k_i}$ is the azimuthal angle of the incident vortex projectile momentum. The vortex TDCS of Eq. (3) can be interpreted as an average of plane wave TDCSs over the azimuthal angle of the incident vortex projectile momentum. These plane wave TDCSs will now be referred to as component TDCSs.

## 2.2 Transition Matrix

For TDCSs with the ionized electron found in the azimuthal plane, past investigations using plane wave collisions showed that the first Born approximation is insufficient to accurately explain experimental results and that elastic scattering of the projectile by the nucleus must be included to account for the double scattering mechanism [1]. Therefore, we implement a distorted wave model, in which the free particle plane wave functions used in the transition matrix of Eq. (3) are replaced by numerical Hartree-Fock distorted wave functions, as described below. We now refer to the plane wave transition matrix in Eq. (3) and its corresponding plane wave TDCS as the non-vortex transition matrix and non-vortex TDCS.

Because the vortex TDCS is expressed as an average over the non-vortex TDCS, we expect that any physical effects that are unimportant in the non-vortex TDCS will also be unimportant in the vortex TDCS and can be safely neglected here. Specifically, it has been shown that for non-vortex projectiles, the inclusion of target electron correlation effects is unimportant for calculation of single ionization TDCSs. Therefore, a single active electron model is sufficient [22–27].

In the non-vortex transition matrix, the initial state wave function is a product of the target helium atom wave function $\Phi(\vec{r}_2)$ and the incident projectile distorted wave $\chi_{\vec{k}_i}(\vec{r}_1)$

$$\Psi_i = \chi_{\vec{k}_i}(\vec{r}_1)\Phi(\vec{r}_2). \tag{4}$$

The atomic electron wave function is given by the one-electron wave function of [28] and the incident non-vortex projectile is given by a numeric Hartree-Fock distorted wave including exchange distortion [29,30].

The final state wave function is a product of the ionized electron wave function $\chi_{\vec{k}_e}(\vec{r}_2)$, the scattered projectile wave function $\chi_{\vec{k}_f}(\vec{r}_1)$, and the post-collision Coulomb interaction $M_{ee}$ [31]

$$\Psi_f = \chi_{\vec{k}_f}(\vec{r}_1)\chi_{\vec{k}_e}(\vec{r}_2)M_{ee}. \tag{5}$$

Both final state outgoing free electron wave functions are modeled as a non-vortex numeric Hartree-Fock wave functions with exchange distortion. The final state ejected electron wave function is orthogonalized to the initial state target helium wave function through Gram-Schmidt orthogonalization.

The post-collision Coulomb repulsion between the two outgoing final state electrons is included through the use of the Ward-Macek factor [31]

$$M_{ee} = N_{ee} \left| {}_1F_1\left(\frac{i}{2k_{fe}}, 1, -2ik_{fe}r_{ave}\right) \right|, \qquad (6)$$

where

$$N_{ee} = \sqrt{\frac{\pi}{k_{fe}\left(e^{\frac{\pi}{k_{fe}}}-1\right)}}. \qquad (7)$$

The relative momentum is $k_{fe} = \frac{1}{2}|\vec{k}_f - \vec{k}_e|$ and the average coordinate $r_{ave} = \frac{\pi^2}{16\epsilon}\left(1 + \frac{0.627}{\pi}\sqrt{\epsilon}\ln\epsilon\right)^2$, where $\epsilon = (k_f^2 + k_e^2)/2$ is the total energy of the two outgoing electrons.

The perturbation is given by

$$V = -\frac{1}{r_1} + \frac{1}{r_{12}} - U, \qquad (8)$$

where the projectile-nuclear distance is $r_1$, the projectile-target electron distance is $r_{12}$, and $U$ is the spherically symmetric distorting potential used to calculate $\chi_{\vec{k}_i}(\vec{r}_1)$. Due to orthogonality of the initial and final state atomic electron wave functions, the projectile-nuclear term's contribution to the transition matrix is zero and is therefore not included in our numerical calculations.

## 2.3 Bessel Wave Function

While the calculation of the vortex transition matrix for Bessel projectiles naturally includes the Bessel wave function, it is not explicitly used in the numerical calculation of the TDCS because Eq. (3) requires only the use of the non-vortex transition matrix [8,10]. Nonetheless, it is useful to discuss some important properties of the incident projectile Bessel wave function, which is given in cylindrical coordinates $(\rho_1, \varphi_1, z_1)$ for $\vec{b} = 0$ by

$$\chi_{\vec{k}_i}(\vec{r}_1) = \frac{e^{il\varphi_1}}{2\pi} J_l(k_{i\perp}\rho_1) e^{ik_{iz}z_1}. \tag{9}$$

The vortex projectile's transverse and longitudinal momenta are $k_{i\perp}$ and $k_{iz}$, respectively, and $l$ is the projectile's quantized orbital angular momentum, also known as the topological charge. The beam's opening angle $\alpha$ is defined through the ratio of projectile transverse and longitudinal momenta

$$\tan\alpha = \frac{k_{i\perp}}{k_{iz}}. \tag{10}$$

Averaging the TDCS over impact parameters washes out any angular momentum information of the projectile, and the TDCS of Eq. (3) only depends on the kinematical parameters of the projectile through the momentum transfer and the vortex opening angle.

As Eq. (9) shows, the Bessel wave function has non-zero transverse momentum. However, the transverse momentum vector is not well-defined; only its magnitude is specified, not its azimuthal angle $\phi_{k_i}$. This results in uncertainty in the incident momentum vector, which in turn leads to uncertainty in the momentum transfer and ionized electron momentum vectors. As shown below, if this uncertainty is large enough, it results in qualitative changes to the shape of the TDCS, such as a broadening of the binary peak.

In the calculation of the transition matrix, it is important to write the momentum transfer vector $\vec{q}$ in terms of its components, explicitly including the vortex opening angle and projectile momentum azimuthal angle. In the coordinate system used here (Fig. 1), the momentum transfer vector components are given by

$$q_x = k_i \sin \alpha \cos \phi_{k_i} - k_f \sin \theta_s \tag{11a}$$

$$q_y = k_i \sin \alpha \sin \phi_{k_i} \tag{11b}$$

$$q_z = k_i \cos \alpha - k_f \cos \theta_s, \tag{11c}$$

where $\theta_s = 90°$ is the projectile scattering angle.

## 2.4 Azimuthal Plane Binary Collisions

A qualitative description of a single binary collision between the projectile and the target atomic electron can be achieved through classical momentum conservation. Assuming an infinitely massive nucleus, momentum conservation for a collision between a projectile and atomic electron gives

$$\vec{k}_i + \vec{k}_{eb} = \vec{k}_f + \vec{k}_e, \tag{12}$$

where $\vec{k}_{eb}$ is the initial state momentum of the bound atomic electron. Separating this equation into transverse and longitudinal components yields

$$\vec{k}_{i\perp} + \vec{k}_{eb\perp} = \vec{k}_{f\perp} + \vec{k}_{e\perp} \tag{13a}$$

$$\vec{k}_{iz} + \vec{k}_{ebz} = \vec{k}_{fz} + \vec{k}_{ez}. \tag{13b}$$

For azimuthal plane ionization, $\vec{k}_{fz} = \vec{k}_{ez} = 0$, and Eq. (13b) implies that the atomic electron's longitudinal momentum must be equal and opposite to the incident projectile's longitudinal momentum

$$\vec{k}_{ebz} = -\vec{k}_{iz}. \tag{14}$$

To determine where the ionized electron is most likely to be ejected, Eq. (13a) can be rewritten as

$$\vec{k}_{e\perp} = \vec{k}_{i\perp} - \vec{k}_{f\perp} + \vec{k}_{eb\perp}. \tag{15}$$

The bound atomic electron's momentum cannot be known explicitly, but its momentum distribution is sharply peaked around $|\vec{k}_{eb}| = 0$, and we can assume that $\vec{k}_{eb} \approx 0$, yielding

$$\vec{k}_{e\perp} \approx \vec{k}_{i\perp} - \vec{k}_{f\perp} \equiv \vec{q}_\perp, \tag{16}$$

where $\vec{q}_\perp$ is the transverse momentum transfer vector.

For non-vortex projectiles, the incident momentum is purely longitudinal, such that $\vec{k}_{i\perp} = 0$ and therefore the ionized electron is most likely to have its transverse momentum equal and opposite to the scattered projectile $\vec{k}_{e\perp} \approx -\vec{k}_{f\perp}$. This mechanism results in the binary peak observed in the non-vortex azimuthal plane TDCS at $\varphi_e = 180°$, i.e. both electrons leave the collision back-to-back.

For vortex projectiles, the incident projectile transverse momentum is non-zero and in general Eq. (16) cannot be simplified further. However, for small opening angles, the incident projectile transverse momentum is small and $\vec{k}_{i\perp}$ can be neglected. This leads to a similar kinematical situation as for non-vortex projectiles, with the ionized electron's transverse momentum being approximately equal and opposite to the scattered electron's momentum. Thus, we expect to see strong back-to-back emission of the outgoing electrons, as was the case for non-vortex projectiles.

For large opening angles, $\vec{k}_{i\perp}$ cannot be neglected. As discussed above, its azimuthal angle $\phi_{k_i}$ is unspecified and spans the range from 0 to $2\pi$. Therefore, the incident projectile's transverse momentum can lie anywhere in the azimuthal plane. Equation (16) then implies that the ionized electron's transverse momentum can also lie anywhere in the azimuthal plane. In other words, the inherent uncertainty in $\vec{k}_{i\perp}$ leads to an inherent uncertainty in $\vec{k}_{e\perp}$, and we expect this uncertainty to lead to changes in the vortex TDCS compared to the non-vortex TDCS for large opening angles.

To more concretely demonstrate the uncertainty in the ionized electron's transverse momentum, Fig. 2 shows the azimuthal angle of the transverse momentum transfer $\varphi_q$ as a function of the incident projectile azimuthal angle $\phi_{k_i}$. From Eq. (16), we expect $\varphi_q$ to be approximately equal to the ionized electron's transverse momentum azimuthal angle $\varphi_e$, and Fig. 2 therefore provides an estimate of the variation in ionized electron azimuthal angle for binary collisions. For vortex projectiles with small opening angles, the range of momentum transfer azimuthal angles is localized around $\varphi_q = 180°$ (the non-vortex momentum transfer azimuthal angle). Therefore, it's expected that the ionized electron will be found primarily near $\varphi_e = 180°$. However, for vortex projectiles with large opening angles, $\varphi_q$ spans the entire angular range between 0 and $2\pi$, and the ionized electron can be found anywhere in the azimuthal plane.

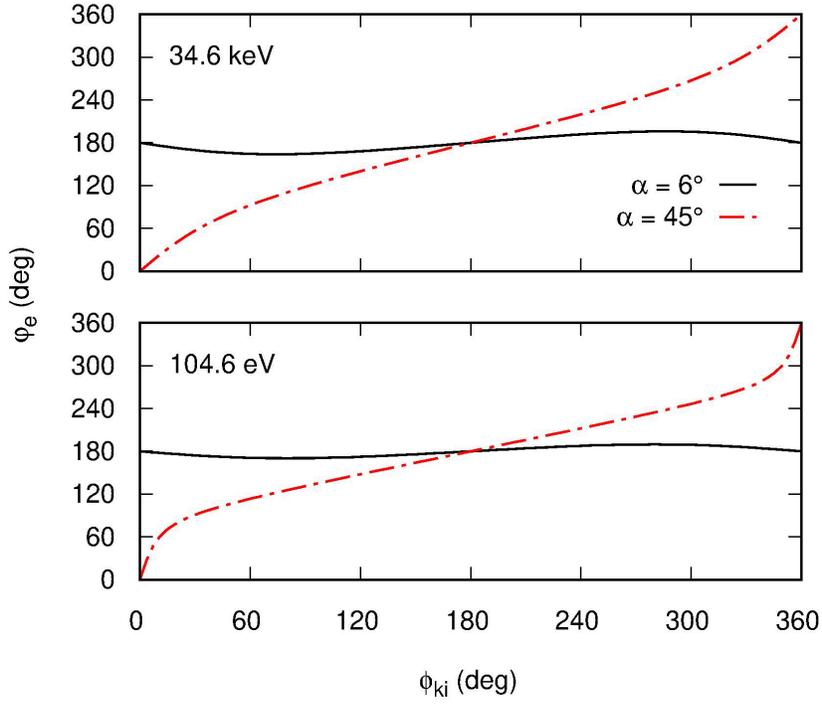

Figure 2 Azimuthal angle of the vortex momentum transfer vectors $\varphi_q$ as a function of incident projectile azimuthal angle $\phi_{k_i}$ for symmetric energy sharing ionization of helium with both final state electrons found in the azimuthal plane. The incident projectile energies are 34.6 eV (top) and 104.6 eV (bottom). Vortex opening angles are $\alpha = 6°$ (black solid line) and $\alpha = 45°$ (red dash-dotted line).

## 3. Results

As discussed above, for non-vortex symmetric energy azimuthal plane ionization, the dominant scattering mechanism at low ionized electron energies is a binary collision identified by a single peak at $\varphi_e = 180°$ in the TDCS due to momentum conservation. However, for larger ejected electron energies, a double scattering mechanism results in additional peaks in the TDCS near $\varphi_e = 90°$ and $270°$. In the double scattering scenario, the projectile electron is first elastically scattered by the nucleus into the azimuthal plane. It then collides with the atomic electron, resulting in both electrons leaving the collision with a relative angle of 90° [1,2]. Note that due to the kinematical symmetry about the scattering plane, the azimuthal plane TDCS are symmetric about $\varphi_e = 180°$, such that the mechanics of electron emission at $\varphi_e = 90°$ or $270°$ are identical.

Because vortex projectiles possess non-zero transverse momentum, we expect that the qualitative features of the azimuthal plane TDCS will differ from those of non-vortex projectiles, and that the simple classical descriptions of the double and single scattering mechanisms may be altered or no longer apply. In particular, for vortex projectiles, ionization into the azimuthal plane is *not* forbidden for a binary collision and can therefore occur with either a single or double collision mechanism. Additionally, the inherent uncertainty in the incident projectile momentum causes uncertainty in the momentum transfer vector, which can significantly alter the shape of the TDCS[9].

### 3.1 Binary Collisions

The top row of Fig. 3 shows the TDCS for ionization into the azimuthal plane for vortex opening angles $0° \leq \alpha \leq 45°$ and an incident energy of 34.6 eV. The left column shows results when a distorted wave is used for the incident projectile and the right column shows results when the incident projectile is treated as a plane wave. The use of a distorted wave for the incident projectile includes elastic scattering from the nucleus and therefore the double scattering mechanism, while these effects are absent for an incident projectile plane wave. At this energy, the non-vortex ionization mechanism is predominantly a binary collision, which is identified by the single peak in the azimuthal plane TDCS at $\varphi_e = 180°$ (Fig. 3a and b, solid black curves, $\alpha = 0$). The similarity between the distorted wave and plane wave non-vortex TDCSs demonstrates that any contribution from the double scattering mechanism is negligible, confirming prior results [1].

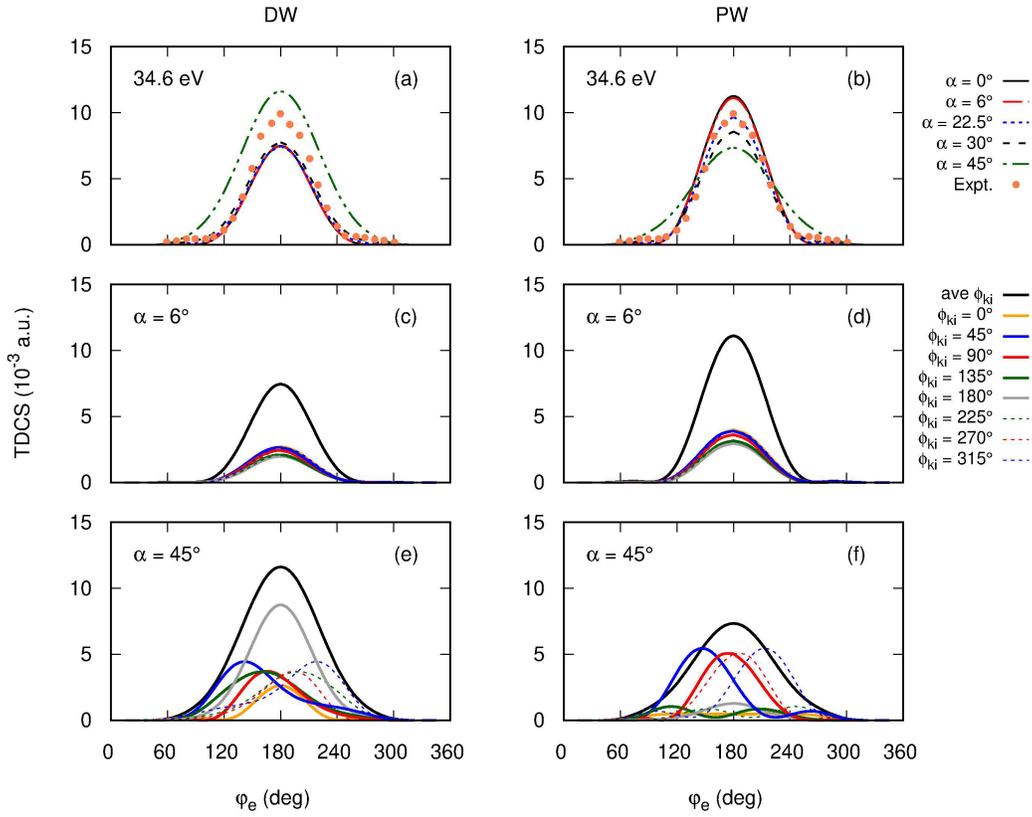

Figure 3 Average and component TDCSs of Eq. (3) for vortex (6° ≤ α ≤ 45°) and non-vortex (α = 0°) ionization into the azimuthal plane with incident projectile energy 34.6 eV. The final state electrons have equal energy (5 eV). The left column (a,c,e) shows results for a distorted wave treatment of the incident projectile and the right column (b,d,f) shows results for a plane wave treatment of the incident projectile. Experimental data are from [2] for non-vortex projectiles and have been normalized to the distorted wave calculation for $\alpha = 0$, $E_i = 104.6$ eV (Fig 4a). Top row: average TDCSs for the opening angles listed in the legend to the right of (b). Middle and bottom rows: component TDCSs for incident projectile azimuthal angles shown in the legend to the right of (d). The solid black curves ('ave $\phi_{k_i}$') in rows 2 and 3 are identical to the red dash-dot and green dash-double dot curves of row 1, respectively.

In contrast to non-vortex projectiles, the different incident projectile momentum vectors of vortex projectiles enable the ejection of an electron at any angle into the azimuthal plane through a single collision. However, the similarity in shape of the vortex and non-vortex TDCS at 34.6 eV indicates that these out-of-plane binary collisions have a minimal effect on the vortex TDCS (Figs. 3a and b), as predicted by the analysis of Section 2.4. Minor changes are observed between the vortex and non-vortex TDCSs at 34.6 eV, including a broadening of the binary peak and an increase (DW) or decrease (PW) in magnitude relative to the non-vortex TDCS for the largest opening angle $\alpha = 45^0$. These changes are a result of the uncertainty in the ionized electron momentum and can be understood by considering the kinematics of binary collisions for vortex projectiles.

Recall that vortex projectiles do not have a single momentum transfer vector, but rather a cone of possible momentum transfer vectors that results from writing the incident vortex projectile as a superposition of tilted plane waves [18]. This leads to a natural uncertainty in the momentum transfer vector for vortex collisions. As shown in Fig. 2, vortex projectiles with small opening angles have a small uncertainty in the momentum transfer azimuthal angle, with most momentum transfer azimuthal angles localized around the non-vortex momentum transfer azimuthal angle ($\varphi_q = 180°$). The nearly identical vortex and non-vortex TDCSs of Fig. 3 confirms our expectation that this limited uncertainty in the momentum transfer vector has little effect on the azimuthal plane TDCS. However, for large opening angles, Fig. 2 shows that vortex projectiles have a large uncertainty in the momentum transfer azimuthal angle, and this results in the broadening of the binary peak in the vortex TDCSs shown in Fig. 3a and b.

Additional insight into the role of the incident projectile's momentum uncertainty can be gained by using the vortex TDCS expression of Eq. (3), which shows that the vortex TDCS can be written as a sum over individual non-vortex TDCSs, (i.e. component TDCSs). Each component TDCS corresponds to a non-vortex TDCS with a unique momentum transfer vector with unique azimuthal angle $\phi_{k_i}$. The binary peak location of the component TDCSs are determined by the momentum transfer direction and can therefore provide information about the mechanics of vortex ionization.

The bottom two rows of Fig. 3 show the component TDCSs for select $\phi_{k_i}$ values. For the smallest opening angle $\alpha = 6°$, the component TDCSs remain localized near $\varphi_e = 180°$, as expected from the localized $\varphi_q$ values shown in Fig. 3. Each component TDCS is nearly identical, indicating that the small amount of incident projectile momentum has a negligible effect on the TDCS. At the larger opening angle $\alpha = 45°$, significant differences are observed between the component TDCSs, and it is clear that the incident projectile's momentum alters the vortex TDCS. Some of the component TDCSs have their binary peaks shifted away from $\varphi_e = 180°$, which is a direct consequence of the incident projectile's transverse momentum. These shifts also break the symmetry of each component TDCS about $\varphi_e = 180°$ and combine to cause the broadening of the binary peak. Qualitatively, the shifts of the component TDCSs are similar for both the plane wave and distorted wave treatments of the projectile, confirming that they are primarily a result of single collisions.

The changes in magnitude observed for the $\alpha = 45°$ binary peaks are different for plane wave and distorted wave incident projectiles. This indicates that elastic scattering from the nucleus does affect the vortex TDCS. In particular, the $\phi_{k_i} = 180°$ component TDCS (gray lines in Fig. 3e and f) is much larger for the distorted wave projectile. At this azimuthal angle, all momenta lie in the scattering plane, and the incident projectile momentum has a large negative x-component. In order for the projectile to be scattered along the positive x-axis, a large change in its momentum is required, which is unlikely with a binary collision. However, if the projectile is able to first elastically scatter from the nucleus such that its momentum vector then has a positive x-component, a smaller change in momentum is required to send it out along the x-axis. This would result in the enhanced component TDCS for distorted wave projectiles at $\phi_{k_i} = 180°$ compared to the plane wave component TDCS.

Unfortunately, no experimental results exist for ionization by electron vortex projectiles. However, the results above present a possible test for determining the scattering mechanism that leads to ionization at low energy. For non-vortex projectiles, it is established that elastic scattering of the projectile from the nucleus is unimportant at low energy. However, our results predict that the vortex TDCS will be larger than the non-vortex TDCS if double scattering is important. Therefore, a direct comparison of vortex and non-vortex TDCSs can indicate whether double scattering is relevant in vortex ionization.

### 3.2 Double Scattering

Figure 4 shows the vortex and non-vortex TDCS, as well as component TDCSs, for an incident projectile energy of 104.6 eV. For non-vortex projectiles at the larger energy, the differences between plane wave and distorted wave treatments of the projectile become more apparent due to the increased importance of the double scattering mechanism. In particular, inclusion of the double scattering mechanism is required to accurately predict the experimental peaks at $\varphi_e = 90°$ and $270°$.

The top row of Fig. 4 shows that vortex projectiles can significantly alter the shape and magnitude of the TDCS, particularly for large opening angles where the incident projectile's momentum is significantly different than its non-vortex counterpart. This is true regardless of whether the projectile is treated as a plane wave or distorted wave. At the small opening angle, only minor differences are observed between the vortex and non-vortex TDCS. This is again likely a result of the small transverse momentum of the incident projectile, which results in only minor changes to the momentum transfer vector and small uncertainties in its azimuthal angle. For the largest opening angle ($\alpha = 45°$), a significant enchancement in the TDCS is observed for $60° \leq \varphi_e \leq 120°$ and $240° \leq \varphi_e \leq 300°$ at the higher energies. This enhancement occurs for both plane wave and distorted wave projectiles, indicating that it is in part due to binary collisions and that double scattering is not required to emit electrons out of the scattering plane. However, the plane wave and distorted wave models show clear differences for vortex TDCSs, indicating that the double scattering mechanism does influence the TDCS shape and magnitude. For distorted wave projectiles, peaks at $\varphi_e = 90°$ and $270°$ are present for all opening angles and a minimum is observed at $\varphi_e = 180°$. This indicates that back-to-back emission is still present for vortex projectiles, but less likely than perpendicular emission. As in the case of non-vortex projectiles, different shapes of the plane wave and distorted wave azimuthal TDCSs point to their usefulness in identifying single vs. double scattering mechanisms in the ionization process.

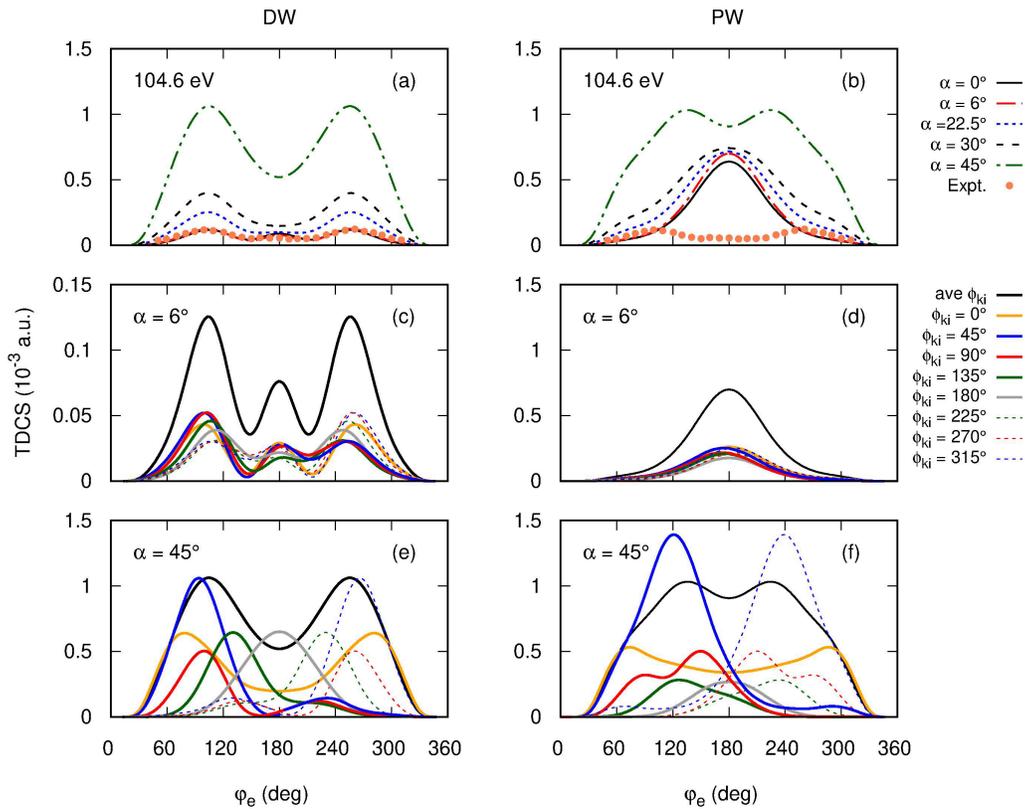

Figure 4 Same as Fig. 3, but with an incident projectile energy of 104.6 eV.

Insight into the features of the vortex TDCS can again be gained by examining the component TDCSs. As was the case for the lower energy, the component TDCSs for $\alpha = 6°$ are all very similar and have the same shape as the non-vortex TDCS. For $\alpha = 45°$, the shift of the component TDCS peaks away from $\varphi_e = 180°$ is greater than at the lower energy and results in not just a broadening of the binary peak, but a splitting of it (see Fig. 4e and f). These shifts are again present for plane wave and distorted wave projectiles, indicating that the splitting of the 'binary' peak is at least partly due to single collisions and again confirming that double scattering is not required to eject electrons into the azimuthal plane for vortex projectiles. This is in contrast to the non-vortex TDCSs, in which double scattering is required to explain the peaks not at $\varphi_e = 180°$.

Similar to the lower energy, the relative magnitudes of the component TDCSs are different for plane wave and distorted wave projectiles. The inclusion of elastic scattering from the nucleus with the distorted wave model increases the magnitude of the $\phi_{k_i} = 180°$ TDCS (gray line in Fig. 4e), indicating again that an initial projectile scattering from the nucleus may reorient the projectile momentum in a direction that enhances back-to-back emission for this particular projectile momentum azimuthal angle. In this case, comparison with experiment would provide evidence as to the importance of back-to-back emission and the double scattering mechanism. The relative depth of the minimum at $\varphi_e = 180°$ can serve as a test of the importance these ionization mechanisms.

## 4. Conclusions

We have presented TDCSs for ionization by electron vortex projectiles in which both final state electrons are found in the azimuthal plane with equal energy. This collision geometry was chosen to highlight single and double scattering mechanisms as incident projectile energy was varied. We compared TDCSs for vortex and non-vortex projectiles and used plane wave and distorted wave treatments of the projectile to gain insight into the scattering dynamics.

For non-vortex collisions, it is well-known that single collisions are the primary cause of ionization at low projectile energy, with both electrons being emitted in a back-to-back geometry. Our results showed that this is also true for vortex collisions, although some double scattering may lead to an enhancement of the binary peak in which both final state electrons leave the collision back-to-back. For low energy projectiles, very little qualitative difference was observed between the vortex and non-vortex TDCSs, with only small changes in magnitude and broadening of the binary peak observed for projectiles with the largest opening angle.

At higher projectile energy, non-vortex collisions are dominated by a double scattering mechanism that results in the final state electrons being ejected perpendicular to each other. For vortex projectiles, a double scattering mechanism is not required to achieve perpendicular emission of the electrons, however our results showed that perpendicular emission was enhanced by inclusion of the double scattering mechanism. At higher projectile energy, more significant changes in magnitude were observed in the TDCSs for vortex and non-vortex projectiles. In particular, the TDCS for the largest vortex opening angle was an order of magnitude larger than that of the non-vortex TDCS.

In our model, the vortex TDCS was written as an average over the non-vortex TDCSs for different projectile momentum azimuthal angles. This feature allowed us to examine the role of the component TDCSs. We showed that for both low and high energy vortex projectiles, the double scattering mechanism enhances back-to-back emission for projectiles with the largest momentum transfer values ($\phi_{k_i} = 180°$). We also showed that the uncertainty in the incident projectile momentum leads to an uncertainty in the ionized electron momentum, which results in a broadening and/or splitting of the binary peak.

Our results offer several predictions that can be directly tested by comparison with experiment. In particular, if the vortex binary peak at low energy is increased relative to the non-vortex binary peak, then double scattering is important. Also, if double scattering is important for vortex projectiles at high projectile energy, then two distinct peaks should be observed in the TDCS. The relative magnitude of the minimum at $\varphi_e = 180°$ can provide insight into the role of back-to-back emission. Unfortunately, no experimental data is available for electron vortex collisions with atoms or molecules, but we hope that this work prompts interest from our experimental colleagues.

**Acknowledgements**

We gratefully acknowledge the support of the National Science Foundation under Grant No. PHY-1912093.

**References**

[1]   Zhang X, Whelan C T and Walters H R J 1990 Energy sharing (e,2e) collisions-ionisation of helium in the perpendicular plane *J. Phys. B: At. Mol. Opt. Phys.* **23** L173–8
[2]   Murray A J, Woolf M B J and Read F H 1992 Results from symmetric and non-symmetric energy sharing (e, 2e) experiments in the perpendicular plane *J. Phys. B: At. Mol. Opt. Phys.* **25** 3021–36
[3]   Bowring N J, Murray A J and Read F H 1997 Near-threshold doubly-symmetric (e, 2e) measurements in helium *J. Phys. B: At. Mol. Opt. Phys.* **30** L671–6
[4]   Rasch J, Whelan C T, Allan R J, Lucey S P and Walters and H R J 1997 Strong interference effects in the triple differential cross section of neutral-atom targets *Phys. Rev. A* **56** 1379–83
[5]   Hawley-Jones T J, Read F H, Cvejanovic S, Hammond P and King G C 1992 Measurements in the perpendicular plane of angular correlations in near-threshold electron impact ionization of helium *J. Phys. B: At. Mol. Opt. Phys.* **25** 2393–408
[6]   Miller F K, Walters H R J and Whelan C T 2015 Energy-sharing $(e,2e)$ collisions: Ionization of the inert gases in the perpendicular plane *Phys. Rev. A* **91** 012706

[7] Murray A J 2015 Ionization differential cross section measurements for H2, D2 and HD at low energies *J. Phys. B: At. Mol. Opt. Phys.* **48** 245203
[8] Harris A L, Plumadore A and Smozhanyk Z 2019 Ionization of hydrogen by electron vortex beam *J. Phys. B: At. Mol. Opt. Phys.* **52** 094001
[9] Plumadore A and Harris A L 2020 Projectile transverse momentum controls emission in electron vortex ionization collisions *J. Phys. B: At. Mol. Opt. Phys.* **53** 205205
[10] Harris A L, Plumadore A and Smozhanyk Z 2020 Corrigendum: Ionization of hydrogen by electron vortex beam (2019 J. Phys. B: At. Mol. Opt. Phys. 52 094001) *J. Phys. B: At. Mol. Opt. Phys.* **53** 109501
[11] Lloyd S M, Babiker M and Yuan J 2012 Interaction of electron vortices and optical vortices with matter and processes of orbital angular momentum exchange *Phys. Rev. A* **86** 023816
[12] Ivanov I P and Serbo V G 2011 Scattering of twisted particles: Extension to wave packets and orbital helicity *Phys. Rev. A* **84** 033804
[13] Ivanov I P and Serbo V G 2011 Erratum: Scattering of twisted particles: Extension to wave packets and orbital helicity [Phys. Rev. A 84, 033804 (2011)] *Phys. Rev. A* **84** 069906
[14] Lloyd S, Babiker M and Yuan J 2012 Quantized Orbital Angular Momentum Transfer and Magnetic Dichroism in the Interaction of Electron Vortices with Matter *Phys. Rev. Lett.* **108** 074802
[15] Afanasev A, Carlson C E and Mukherjee A 2013 Off-axis excitation of hydrogenlike atoms by twisted photons *Phys. Rev. A* **88** 033841
[16] Matula O, Hayrapetyan A G, Serbo V G, Surzhykov A and Fritzsche S 2014 Radiative capture of twisted electrons by bare ions *New J. Phys.* **16** 053024
[17] Van Boxem R, Partoens B and Verbeeck J 2014 Rutherford scattering of electron vortices *Physical Review A* **89**
[18] Van Boxem R, Partoens B and Verbeeck J 2015 Inelastic electron-vortex-beam scattering *Physical Review A* **91** 032703
[19] Serbo V, Ivanov I P, Fritzsche S, Seipt D and Surzhykov A 2015 Scattering of twisted relativistic electrons by atoms *Phys. Rev. A* **92** 012705
[20] Kosheleva V P, Zaytsev V A, Surzhykov A, Shabaev V M and Stöhlker Th 2018 Elastic scattering of twisted electrons by an atomic target: Going beyond the Born approximation *Phys. Rev. A* **98** 022706
[21] Lei C and Dong G 2021 Chirality-dependent scattering of an electron vortex beam by a single atom in a magnetic field *Phys. Rev. A* **103** 032815
[22] Jones S, Madison D H, Franz A and Altick P L 1993 Three-body distorted-wave Born approximation for electron-atom ionization *Phys. Rev. A* **48** R22–5
[23] Berakdar J and Briggs J S 1994 Three-body Coulomb continuum problem *Phys. Rev. Lett.* **72** 3799–802
[24] Srivastava M K and Sharma S 1988 Triple-differential cross sections for the ionization of helium by fast electrons *Phys. Rev. A* **37** 628–31
[25] Bellm S, Lower J, Bartschat K, Guan X, Weflen D, Foster M, Harris A L and Madison D H 2007 Ionization and ionization–excitation of helium to the n = 1 – 4 states of He + by electron impact *Physical Review A* **75**
[26] Dupre C, Lahmam-Bennani A, Duguet A, Mota-Furtado F, O\textquotesingleMahony P F and Cappello C D 1992 (e,2e) triple differential cross sections for the simultaneous ionization and excitation of helium *J. Phys. B: At. Mol. Opt. Phys.* **25** 259–76

[27] Balashov V V and Bodrenko I V 1999 Triple coincidence (e,2egamma) measurements as a \textasciigraveperfect experiment\textquotesingle instrument for ionization-excitation studies *J. Phys. B: At. Mol. Opt. Phys.* **32** L687–92

[28] Byron F W and Joachain C J 1966 Correlation Effects in Atoms. I. Helium *Phys. Rev.* **146** 1–8

[29] Prideaux A, Madison D H and Bartschat K 2005 Exchange distortion and postcollision interaction for intermediate-energy electron-impact ionization of argon *Phys. Rev. A* **72** 032702

[30] Furness J B and McCarthy I E 1973 Semiphenomenological optical model for electron scattering on atoms *J. Phys. B: Atom. Mol. Phys.* **6** 2280–91

[31] Ward S J and Macek J H 1994 Wave functions for continuum states of charged fragments *Phys. Rev. A* **49** 1049–56